\documentclass[prd,preprintnumbers,superscriptaddress,nofootinbib,floatfix,twocolumn,notitlepage]{revtex4}
\usepackage[dvips]{graphicx}
\usepackage{dcolumn} 
\usepackage{bm}
\usepackage{epsfig,amsmath,amssymb,verbatim,mathrsfs,array,layout,textcomp,amssymb,latexsym,slashed}
\usepackage{color}
\usepackage{hyperref}
\usepackage[utf8]{inputenc}

\input epsf.tex
\newcommand{\beq}{\begin{eqnarray}}
\newcommand{\eeq}{\end{eqnarray}}

\def\ltap{\ \raise.3ex\hbox{$<$\kern-.75em\lower1ex\hbox{$\sim$}}\ }
\def\gtap{\ \raise.3ex\hbox{$>$\kern-.75em\lower1ex\hbox{$\sim$}}\ }

\def\eg{{\it e.g.}}

\def\be{\begin{equation}}
\def\ee{\end{equation}}
\def\bea{\begin{eqnarray}}
\def\eea{\end{eqnarray}}

\definecolor{newred}{rgb}{0.5,0.1,0}
\definecolor{darkgreen}{rgb}{0.0,0.7,0.2}
\definecolor{lightblue}{rgb}{0.0,0.5,1}

\RequirePackage[normalem]{ulem} 
\RequirePackage{color}\definecolor{RED}{rgb}{1,0,0}\definecolor{BLUE}{rgb}{0,0,1} 

\begin{document} 

\title{Probing Atomic Higgs-like Forces at the Precision Frontier}

\author{C\'edric Delaunay}
\email{cedric.delaunay@lapth.cnrs.fr}
\affiliation{LAPTh, Universit\'e Savoie Mont Blanc, CNRS B.P. 110, F-74941 Annecy-le-Vieux, France}
\author{Roee Ozeri}
\email{roee.ozeri@weizmann.ac.il}
\affiliation{Department of Physics of Complex Systems, Weizmann Institute of Science, Rehovot 7610001, Israel}
\author{Gilad Perez}
\email{gilad.perez@weizmann.ac.il}
\affiliation{Department of Particle Physics and Astrophysics, Weizmann Institute of Science, Rehovot 7610001, Israel}
\author{Yotam Soreq}
\email{soreqy@mit.edu}
\affiliation{Center for Theoretical Physics, Massachusetts Institute of Technology, Cambridge, MA  02139, U.S.A.}

\begin{flushright}
\preprint{\scriptsize LAPTH-003/16\vspace*{.1cm}}
\preprint{\scriptsize MIT-CTP/4762\vspace*{.1cm}}
\end{flushright}
\vskip .05in

\begin{abstract}
We propose a novel approach to probe new fundamental interactions using isotope shift spectroscopy in atomic clock transitions. As concrete toy example we focus on the Higgs boson couplings to the building blocks of matter: the electron and the up and down quarks. We show that the attractive Higgs force between nuclei and their bound electrons, that is poorly constrained, might induce effects that are larger than the current experimental sensitivities. 
More generically, we discuss how new interactions between the electron and the neutrons, mediated via light new degrees of freedom,  may lead to measurable non-linearities in a King plot comparison between isotope shifts of two different transitions. Given state-of-the-art accuracy in frequency comparison, isotope shifts have the potential of being measured with sub-Hz accuracy, thus potentially enabling the improvement of current limits on new fundamental interactions.  Candidate atomic system for this measurement require two different clock transitions and four zero nuclear spin isotopes. We identify several systems that satisfy this requirement and also briefly discuss existing measurements. We consider the size of the effect related to the Higgs force and the requirements for it to produce an observable signal. 
\end{abstract}

\maketitle

{\bf Introduction}
The Standard Model~(SM) of elementary particles and fundamental interactions is one of the most successful scientific theories. Its last piece, the Higgs boson, was recently observed by the Large Hadron Collider~(LHC) experiments~\cite{ATLASdisco,CMSdisco}, a discovery of uttermost importance that led to the awarding of the 2013 Nobel Prize in Physics. Yet, the SM cannot be a complete description of nature. It does not possess a viable dark matter candidate, neutrino masses and mixings are unaccounted for, and it cannot explain the matter-antimatter asymmetry of our Universe. Furthermore, the masses of the charged fermions (quarks and leptons) exhibit a strong hierarchy, leading to the celebrated ``flavor puzzle". 

Within the SM, every fermion mass $m_f$ is induced by the product $y_f\times v$, where $y_f$ denotes the fermion coupling to the Higgs boson, which corresponds to the strength of the Higgs force felt by the fermion $f$, and $v=(\sqrt{2}G_{\rm F})^{-1/2}\approx246\,$GeV is the Higgs vacuum expectation value~(VEV). Hence, the observed hierarchy of masses is a result of the {\it assumed} hierarchy in $y_f$, leading to a unique construct with
\begin{equation}\label{ySM}
y_f^{\rm SM} = {m_f\over v}\,.
\end{equation}
The Higgs discovery leads us to ask: is this new particle indeed the SM Higgs? It is possible that some of its properties point to physics beyond the SM. An important new physics test is related to the Higgs boson couplings to the building blocks of matter: the electron  and the up and down quarks, the proton's and neutron's ``valence'' quarks. We are currently rather ignorant regarding these, and within the SM these couplings (evaluated at the Higgs mass scale) are very small~\cite{Xing:2011aa},
\begin{equation}
y_{e,u,d}^{\rm SM}(m_h)\simeq 2.0\times 10^{-6}\,, \ 5.4\times 10^{-6}\,, \ 1.1\times 10^{-5}\,\,.
\end{equation} 
In fact, it is possible that the strength of these Higgs-to-light-fermion interactions is far stronger than the above prediction~\cite{Giudice:2008uua}, or that the light fermion masses are not due to the Higgs mechanism, resulting in much smaller couplings~\cite{Ghosh:2015gpa}. Furthermore, additional light degrees of freedom that are associated with the breaking of flavor symmetries and might even address the little hierarchy problem~\cite{Gupta:2015uea} may lead to a new scalar force with a larger coupling to the lighter charge fermions~\cite{Flacke:2016szy}. These cases lead to an alternative understanding of the flavor puzzle and to the establishment of new physics~\cite{Frugiuele:2016rii}.

LHC Higgs data only directly probes the light quark couplings through the total Higgs width constraint $\Gamma_h\leq1.7\,$GeV~\cite{Khachatryan:2014jba}. This translates into weak bounds of $y_{u,d,s,c}\lesssim0.3$~\cite{Perez:2015aoa,Zhou:2015wra}, which is $\mathcal{O}(10^4)$ larger than the SM values for $u$ and $d$ quarks. 
Global fits also indirectly constrain the light quark couplings, yet with additional assumptions on the production of the Higgs boson. Currently available LHC Higgs data together with electroweak~(EW) precision tests yield a stronger bound of $y_{u,d,s,c}\lesssim1.6\times10^{-2}$~\cite{charmingtheHiggs,Kagan:2014ila,Perez:2015aoa}. 
The electron coupling, on the other hand, is best probed through the direct search at the LHC of the $h\to e^+e^-$ decay, giving $y_e\lesssim 1.3\times10^{-3}$~\cite{Aad:2014xva,Khachatryan:2014aep}, see also~\cite{Altmannshofer:2015qra}. The above bounds may be improved at next LHC runs. While direct bounds are not expected to significantly improve due to limitations in the detector resolution, indirect bounds from global fits could reach $y_{u,d,s,c}\lesssim 5\times10^{-3}$ at the high luminosity stage~\cite{Perez:2015lra}. The potential reach of the exclusive Higgs decays~\cite{Bodwin:2013gca, Kagan:2014ila,Koenig:2015pha} to bound its couplings to constituent quarks is expected to be even weaker~\cite{Perez:2015lra}. 
Meanwhile, the direct bound on the electron coupling could improve by an order of magnitude~\cite{Altmannshofer:2015qra}. 

An alternative approach to the above experimental program at the energy frontier relies on low-nergy precision measurements, for instance via atomic physics. Frequency measurements of narrow optical clock transitions in heavy atoms recently reached an unprecedented accuracy of $\mathcal{O}(10^{-18})$~\cite{JunYe}. This remarkable level of precision paves the way for new tests of the existence of physics beyond the SM. Applications of atomic clock transitions have already been proposed in order to probe possible time-variation of fundamental constants~\cite{rosenband2008frequency,huntemann2014improved,Godun:2014naa}, and the existence of cosmological relics in the form of  topological defects~\cite{Derevianko:2013oaa} or new ultralight particles~\cite{Arvanitaki:2014faa} possibly associated with dark matter. 
We argue in this letter that sub-Hz precision measurements of isotope shifts in alkali or rare-earth atoms can potentially probe physics related to the origin of charged fermion masses . In this work we focus mostly on the physics of heavy force mediator. A discussion related to light mediators is to be reported elsewhere~\cite{Berengut:2017zuo}.\\

{\bf Higgs force in atoms}
Higgs boson exchange between a nucleus of mass number $A$ and one of its bound electrons induces an attractive potential of Yukawa type, see for example~\cite{Haber:1978jt}, 
\beq\label{VHiggs}
V_{\rm Higgs}(r)=-\frac{y_e y_A}{4\pi}\frac{e^{-r m_h }}{r}\,.
\eeq
$m_h\approx 125\,$GeV is the mass of the physical Higgs boson~\cite{Higgsmass} and $y_A=(A-Z)y_n+Z y_p$ is the effective nuclear coupling; $Z$ is the atomic number and $y_{n,p}$ are respectively the neutron and proton couplings. In terms of fundamental quark couplings (evaluated at the Higgs mass scale), they read~\cite{Shifman:1978zn,Belanger:2008sj,Junnarkar:2013ac,microOmega}
\beq \label{eq:Ynpeff}
\begin{matrix}
	y_n &\simeq&  7.7y_u + 9.4 y_d + 0.75 y_s +2.6\times 10^{-4}c_g \, , \\
	y_p &\simeq&  11y_u + 6.5 y_d + 0.75 y_s +2.6\times 10^{-4}c_g \, , 
\end{matrix}
\eeq 
where $c_g=4.0\times 10^2y_c+88 y_b + 1.5y_t +\delta c_{g}$ is the effective coupling to gluons which includes the $c,b,t$ contributions as well as a possible new physics contribution $\delta c_g$. 
Moreover, in cases where the Higgs mixes with a light degree of freedom $\phi$, with mass $m_\phi$ and no coupling to SM fermions, the exchange of $\phi$ will induce an additional effective potential proportional to the couplings of the Higgs boson, \eg~\cite{Flacke:2016szy},  
\beq\label{VHiggs}
V_{\rm \phi}(r)=-\frac{y_e y_A}{4\pi}\sin\theta\frac{e^{-r m_\phi }}{r}\,, 
\eeq
where $\theta$ is the mixing angle between $\phi$ and the Higgs boson.

LHC data already indicate that the Higgs boson coupling to top and bottom quarks cannot deviate from the SM prediction by more than a factor few~\cite{HiggsCombine}. Given the direct bound above, the charm quark contributes at most $\approx0.03$ to $y_{n,p}$ which is subdominant to the $u,d,s$ contributions. Additional contributions to the Higgs-to-gluon coupling are also constrained\footnote{Sizable contributions to $\delta c_g$ at the GeV scale could arise, while remaining invisible at the LHC, from a new physics sector which couples to QCD between the weak scale and the QCD scale. Such large contributions would however significantly modify the running of the QCD coupling and are therefore challenged by various precision measurements at low and high energies.}, $\delta c_g\lesssim \mathcal{O}(1)$~\cite{HiggsCombine}. We therefore neglect  $c_g$ in the remainder. 
Within the SM, the $u,d,s$ quark couplings are suppressed by the small fermion masses. Therefore, the heavy quarks dominate in Eq.~\eqref{eq:Ynpeff}, yielding small nucleon couplings, $y_{n,p}^{\rm SM}\sim 10^{-3}$.
However, requiring fundamental quark couplings to saturate the direct LHC constraints, nucleon couplings could reach values as large as $y_{n,p}\sim 3$; while they are limited to $y_{n,p}\lesssim 0.2$ by indirect bounds (see discussion above). Consequently, given the direct bounds on the quark and electron couplings, the strength of the Higgs force in atoms could be enhanced by a factor as large as $10^6$ compared with the SM prediction.\\

{\bf Frequency shifts} We evaluate the Higgs contribution to atomic transition frequencies. Despite the possibly large nuclear Higgs coupling, the range of the Higgs interaction is extremely short, of $\mathcal{O}(m_h^{-1})\sim 10^{-3}\,$fm, and its strength remains much weaker than the dominant Coulomb interaction.
 The Higgs shift in electronic energy levels is then well-described in first-order (time-independent) perturbation theory. For the sake of simplicity, we derive our results using non-relativistic wave functions. In this limit,
\beq\label{EHiggs}
\delta E_{nlm}^{\rm Higgs} = \langle nlm|V_{\rm Higgs}|nlm\rangle\simeq -\frac{y_e y_A}{4\pi m_h^2 }  |\psi(0)|^2 \frac{ \delta_{l,0} }{n^3}\,,
\eeq
where the ket $|nlm\rangle $ is a solution of the Sch\"odinger equation for the unperturbed Coulomb potential, while $n\geq1$ and $0\leq l\leq n-1$ ($-l\leq m\leq l$) are, respectively, the principal and angular momentum quantum numbers. $|\psi(0)|^2/n^3$ is the wave-function-density at the origin ($r=0$) for the electron of interest. In order to obtain $\psi(0)$ we solve for the wave function including the presence of the inner shell electrons (see~\cite{AtomicBook} for more details). We note that the transitions considered below are between partial waves of high angular momentum ($D$- or $F$-waves) with negligibly small overlap with the nucleus and an $S$-wave ($l=0$) ground state. Hence, for this proposal, only $S$-wave energies are effectively shifted by the Higgs force. In the vicinity of the nucleus, the electron is typically in a relativistic regime and electron-electron interactions are important. A proper study of the relativistic and many-body effects in this region involves heavy numerical calculations of  electronic structure, which are beyond the scope of this work. However, relativistic theory for electrons in atoms shows that non-relativistic wave functions  yield a reasonably good estimate for $S$-waves around the nucleus~\cite{AtomicBook,khriplovich1991parity}. For $S$-waves the wave-function-density at the origin can be extracted from hyperfine splitting measurements, if available. For atoms with a single electron above closed shells, a good approximation of $|\psi(0)|^2/n^3$ is obtained by replacing $n$ with an effective principal quantum number $n_*$ that can be infered from fitting the measured binding energy to the Rydberg formula, for example see~\cite{sobelman1979atomic}.

The frenquency shift $\Delta\nu = \Delta(\delta E)/2\pi\hbar$ resulting from the Higgs force for a $n,l\to n',l'$ transition can be conveniently written as 
\beq\label{higgsnushift}
\Delta\nu_{nl\to n'l'}^{\rm Higgs}= 2.6\times  10^2\, {\rm Hz}\times y_e y_A   {|\psi(0)|^2\over 4a_0^{-3}}I_{nn'}^{ll'}
 \,,
\eeq
with $a_0\equiv(\alpha\, m_e)^{-1}$ is the Bohr radius, $\alpha$ is the fine structure constant and $I_{nn'}^{ll'}\equiv( {\delta_{l,0}/ n_*^3}-{\delta_{l',0}/ n_*^{\prime\,3}})$.\\


{\bf Atomic clock transitions}
The most accurate frequency measurements to date have been performed on narrow optical-clock transitions in laser-cooled atoms or ions, where state-of-the-art frequency comparisons are made with relative uncertainty in the $10^{-18}$ range~\cite{NIST,JunYe}. Moreover, various spectroscopic investigations of optical-clock transitions in alkali-like systems are performed with sub-Hz accuracy~\cite{NRC, Innsbruck, PTB}. We argue in the following that, given the current collider bounds, the Higgs-mediated contributions in these atoms are potentially much larger than the experimental sensitivity. 

Consider for instance the optical electric-quadrupole $nS_{1/2}\rightarrow n'D_{5/2}$ transition in $^{88}$Sr$^+\ (Z=38, n=5, n' = 4, n_*\approx2.2)$ or in $^{40}$Ca$^+\ (Z=20, n=4, n' = 3, n_*\approx2.1)$. We use Eq.~\eqref{higgsnushift} to estimate the expected frequency shift.
In this case  $|\psi(0)|^2\simeq 4(1+n_e)^2\,Z/a_0^{3}\,,$ where the density of the valence electron  at the nucleus approximately scales linearly with the nuclear charge $Z$ (and not like $Z^3$) due to the screening of core electrons~\cite{AtomicBook}. 
We have included a factor of $(1+n_e)^2$ relative to the result of~\cite{AtomicBook} to account for the fact that in ions the typical radius of the valence electron is shorter by a factor $\simeq 1+n_e$, where $n_e$ is the ion charge.  
Thus, the Higgs contributions could be as large as  roughly $1\,$kHz and $300\,$Hz, respectively, with saturated bounds on the Higgs couplings. With reported accuracy of these transitions being below $1\,$Hz~\cite{NRC, Innsbruck}, corresponding to a relative accuracy of $\sim10^{-15 }$, the experimental uncertainty on the evaluation of Higgs couplings would be of $ y_ey_{n,p} \lesssim  4\times10^{-6}$, which is stronger than current collider (direct) bounds by a factor of $\sim1000$. 

An even higher sensitivity to Higgs couplings can be obtained in Yb$^+\ (Z=70)$, where the Higgs shift is enhanced by the larger number of nucleons, $A$. A unique benefit of Yb$^+$ is the presence of two narrow transitions in the optical range, namely the electric-quadrupole~(E2) $6S_{1/2}\rightarrow 5D_{3/2}$ and the electric-octupole~(E3) $S_{1/2}(4f^{14}6s)\rightarrow F_{7/2}(4f^{13}6s^2)$ transitions (note that $n_*\approx2.1$ from the ground state). 
Both transitions have also been recently measured with sub-Hz accuracy ($0.36\,$Hz~\cite{PhysRevA.89.023820} and $0.25\,$Hz~\cite{huntemann2014improved}, respectively, see also~\mbox{
\cite{Godun:2014naa}}
), yielding an uncertainty on extracting the Higgs coupling of $y_ey_{n,p} \lesssim 2\times 10^{-7}$. Therefore, from an experimental point of view, the study of Higgs-mediated interactions in laser-cooled atoms seems very promising. On the theory side, the situation is much less promising. Indeed, the effect of many-body electron-electron interactions, along with different contributions that arise from the interaction of the valence electron with the nucleus, are not sufficiently known to be accounted for on the $10^{-15}$ level.\\

{\bf Isotope shifts} An alternative to comparison of absolute frequency measurements to theory would be to scrutinize frequency differences between several isotopes for the optical clock transitions. In principle these isotope shifts (IS) could also be measured with sub-Hz accuracy and their theory predictions are subject to less uncertainties since the total charge $Z$ remains constant. The Higgs contribution to the IS is roughly that of the individual transition frequency times the relative mass change between isotopes. 
For instance, in a frenquency comparison between $^{40}$Ca$^+$ and $^{48}$Ca$^+$, for the optical clock transition above, a change of $\simeq20\%$ in mass leads to a Higgs contribution to the IS of $\sim60\,$Hz  with saturated bounds ($ y_ey_n\simeq0.004$). In a similar comparison between $^{86}$Sr$^+$ and $^{88}$Sr$^+$ ($^{168}$Yb$^+$ and $^{176}$Yb$^+$) a contribution of $ \sim20\,$Hz ($\sim200\,$Hz) is expected. IS in Sr$^+$ were recently measured with a precision of $\sim4\,$kHz~\cite{ChiaveriniClockIS}. Although experimental improvement down to the $1\,$Hz level for this very clock transition is realistic, theoretical calculations are still far from being able to predict the exact IS frequency in these atoms with such precision. In particular, the nuclear charge radius and many-body electron correlations typically result in large uncertainties. For instance, an {\it ab initio} frequency calculation by the authors of Ref.~\cite{ChiaveriniClockIS} disagrees with their measurement by more than $20\%$, with a discrepancy of $\sim100\,$MHz.\\

{\bf Breaking King's linearity}
The IS between $A$ and $A'$ isotopes is usually thought of as arising from two different contributions: a mass shift~(MS) and a field shift~(FS)~\cite{king2013isotope}. The MS is due to the nuclear mass change between the two isotopes. It receives contribution from a change in nuclear recoil (normal MS) and a change in electron-electron correlations (specific MS). Both effects are proportional to the relative mass change  $\mu_{AA'} \equiv 1/m_A - 1/m_{A'}=(A'-A)/(AA')\,$amu$^{-1}$, where ${\rm amu}\approx0.931\,$GeV is the atomic mass unit. The FS, on the other hand, is due to the change in the charge distribution of the nucleus and it is approximately proportional to $\delta\langle  r^2 \rangle_{AA'}$, the difference in the charge distribution variance between the two isotopes. Therefore, the IS for a given transition $i$ is assumed to be of the form 
\begin{equation} \label{eq:ISbasic}
\delta \nu^{AA'}_i \equiv \nu_i^A-\nu_i^{A'}= K_i\,\mu_{AA'} + F_i\delta\langle  r^2 \rangle_{AA'},
\end{equation}
where $K_i$ and $F_i$ are, respectively, the MS and FS coefficients, that only depend on the transition, not on the isotopes. 
Both the specific MS and the FS pose a serious difficulty in calculating the IS from first principles as the change in nuclear charge radius and the proportionality factors in both cases are non-perturbative quantities. 

A standard way to extract ratios and differences between the proportionality factors above, for two different transitions, and without knowledge of $\delta\langle  r^2 \rangle_{AA'}$, is the King plot~\cite{King:63}. Defining modified IS as $m\delta\nu_{AA'}^i \equiv \delta \nu_{AA'}^i/ \mu_{AA'}$, the change in charge radius between isotopes can be extracted from the IS in a single transition ($i=1$) as $\delta\langle  r^2 \rangle_{AA'}/\mu_{AA'} = (m\delta\nu_{AA'}^1 - K_1)/F_1$ and substituted in the IS expression for a second transition ($i=2$), which yields
\beq \label{eq:Kingbasic}
m\delta\nu^2_{AA'} = F_{21} m\delta\nu^1_{AA'} +K_{21}\,,
\eeq 
with $K_{21}\equiv(K_2-F_{21}K_1)$ and $F_{21}\equiv F_2/F_1$. A linear relation between the (modified) IS associated with two different transitions is therefore expected. If data are consistent with this linear relation, its slope $F_{21}$ and offset $K_{21}$ can then be extracted by plotting the IS of two transitions against each other for several isotope pairs.

With experimental accuracy below the Hz level, IS measurements will become in principle sensitive to faint weak and Higgs contributions, in the presence of which Eq.~\eqref{eq:ISbasic} becomes 
\beq\label{eq:ISnew}
\delta \nu_{AA'}^i = K_i\,\mu_{AA'} + F_i\delta\langle  r^2 \rangle_{AA'}+H_i(A-A')\,,
\eeq
with $H_i\equiv 2.7\times  10^2\,$Hz $\times \, (1+n_e)^2Z  I_{nn'}^{ll'}(y_ey_n-4.9\times 10^{-3}q_W^n)$ where $q_W^n$ is the weak nuclear charge per neutron. In the SM, $q_W^n= -1$  at tree level. The King relation in Eq.~\eqref{eq:Kingbasic} is in turn modified as
\beq\label{eq:KingCool}
m\delta\nu^2_{AA'} 
&=&	F_{21} m\delta\nu^1_{AA'} +K_{21}  - A A'  H_{21}\,,\label{HnlKing}
\eeq
where we defined $H_{21}\equiv (H_2-F_{21}H_1)\,$amu. Equation~\eqref{HnlKing} shows that the Higgs and weak contributions explicitly break King's linearity law. A couple of comments are in order: 
\begin{itemize} 
\item 
Viewed from the atomic length scale, the finite nuclear size is characterized by a local interaction at the nucleus, like the Higgs and weak forces. Hence, to leading order, $H_i\propto F_i \propto |\psi(0)|^2$, which results in a vanishing $H_{21}$ up to residual effects of the nuclear charge radius over the atomic radius, thus suppressing  the sensitivity to Higgs couplings.\footnote{We thank Krzysztof Pachucki and Maxim Pospelov for bringing this point to our attention.} To our knowledge there is no precise calculation of $H_i$ besides the above non-relativistic estimate, and we parameterize below the possible alignment between the $F_i$ and $H_i$ constants by a factor of $S_{21}\equiv 1-(F_2/F_1)(H_1/H_2)$. 
In Ref.~\cite{Berengut:2017zuo}, it was argued that in the heavy mass limit, 
$H_{21}\propto m^{-3}$\, which implies $S_{21}\sim (a_0\, m)^{-1}\,,$ with $m$ being the mediator mass (for instance the Higgs mass).
\item There is a possibility for nature to accidentally conspire to cancel this non-linearity if $m\delta\nu_{AA'}^i$ are linear functions of $A'$. In this case, the $H_{21}$ term is a mere correction to the slope parameter $F_{21}$ and sensitivity to any effect contributing to $H_i$ is lost.  While the precise isotopic dependence of $m\delta\nu_{AA'}^i$ is straightforward to check directly from data, once available, we note that theory estimates strongly disfavour linear scaling of $m\delta\nu_{AA'}^i$ with $A'$. This is expected because the charge radius of nuclei depends on their shell structure and therefore does not increase monotonically with the number of neutrons; see e.g.~\cite{NerloPomorska:1994pw,Wang:2013zia}.  We thus find these accidental cancellation to be unlikely. 
\end{itemize} 

It is possible therefore that, in the presence of new type of force mediator between the electron and nucleus, the King's law would break. Such an effect may be potentially observed in narrow optical clock transitions. Conversely, as long as IS data remains consistent with the King relation in Eq.~\eqref{eq:Kingbasic}, $H_{21}$ can be bound largely independently of theory uncertainties. Furthermore, with sufficiently good knowledge of the atomic structure, in particular $|\psi(0)|^2$, and of the weak charge per neutron $q_W^n$, the $y_ey_n$ combination of Higgs couplings can be constrained. State-of-the-art many-body simulations already predict the atomic structure of single-valence electron systems below the $1\%$ level~\cite{Porsev:2009pr}.\\ 

{\bf The case of Yb ion}
At least four isotopes are needed in order to probe the third term on the RHS of Eq.~\eqref{eq:KingCool} through a deviation from linearity in a King plot. 
 To this end, an appealing option is to use the
 two narrow optical clock transitions of Yb$^+$, namely the E2 and E3 transitions at 436\,nm and 467\,nm, respectively. In this case, 
\begin{align}
	\label{eq:H21Ybp}
	\frac{H_{21}^{{\rm Yb}^+}}{{\rm Hz\,amu}}\approx  
 	 4\times 10^{3}\,y_e y_nS_{21}\,.
\end{align}
The Higgs force could appear slightly below $20\,$Hz under current constraints. The resulting sensitivity on the Higgs or other form of similar new physics can be estimated as follows. First of all, we assume that the weak contribution is subtracted from $m\delta\nu$'s with sufficient accuracy, and that a King plot constructed from the remainder IS shows a linear behavior. Thus, from Eq.~\eqref{eq:KingCool}, $H_{21}$ is bounded to be smaller than the error on $(m\delta \nu^2_{AA'} - F_{21}m\delta \nu^1_{AA'} - K_{21})/AA'$, which we take to be dominated by the IS measurement uncertainty $\Delta$, yielding~\cite{Berengut:2017zuo}
\begin{align}
	\label{eq:yeyn}
	y_e \, y_n \lesssim \frac{2\times 10^{-3}}{|S_{21}|} \left(\frac{\Delta}{\rm Hz}\right)\left(\frac{17Za_0^{-3}}{|\psi(0)|^2}\right) \left(\frac{8}{A'-A}\right) \,.
\end{align}

As argued above, the reach of the method is suppressed in the limit of $S_{21}\to0$, that is expected to arise for heavy mediators, with masses above the scale that corresponds to the inverse of the nucleus size~\cite{Berengut:2017zuo}. While $F_{21}$ could be extracted directly from the slope of the linear King plot, the ratio of $F_{21}$ and $H_2/H_1$ needs to be calculated. We note that despite the fact that a $6S$ electron is active in both transitions, one may expect that $1-F_2/F_1\sim\mathcal{O}(1)$. The reason stems from the significantly different influence of core electrons between the E2 transition, where the $4F$ shell is complete, and the E3 one, where it is missing one electron.

Combining Eq.~\eqref{eq:yeyn} with Eq.~\eqref{eq:Ynpeff}, one obtains a sensitivity to the fundamental Higgs-to-light-quark couplings of
\begin{align}
	y_u + 1.2y_d + 0.10y_s \lesssim \frac{0.2}{|S_{21}|} \left(\frac{1.3\times10^{-3}}{y_e} \right)\left(\frac{\Delta}{\rm Hz}\right) \,,
\end{align}
neglecting the subdominant heavy quarks contribution.

Typical IS for clock transitions in Yb$^+$ are in the GHz range, for example~\cite{PhysRevA.85.012502}, while the experimental sensitivity is of  $\mathcal{O}({\rm 0.1 Hz})$. Thus it is an important question to understand what is the expected size of the residual contributions from QED and the strong force, which were neglected in Eq.~\eqref{eq:Kingbasic}. In particular whether these contributions are sufficiently suppressed and at most $\mathcal{O}(10^{-9})$ relative to the leading terms. A parametric argument, in the non-relativistic limit, shows that non-linearities in a King plot induced by the nuclear effects are at least $10^{-14}$ and $10^{-10}$ smaller than the dominant IS contributions from FS and MS, respectively. We give below the general lines of the argument. First of all, observe that the IS is controlled by two small parameters: the difference of the electron reduced masses divided by their sum, $\approx (m_e/2m_p)(1/A-1/A') \simeq (A'-A)\varepsilon_{\mu}$ with $\varepsilon_{\mu}\sim10^{-8}$, and the change in nuclear rms charge radius divided by the ion size, $\langle \delta r^2 \rangle_{AA'}(\alpha m_e)^2\sim (A'-A) \varepsilon_r$ with $\varepsilon_r\sim10^{-11}$. The MS and FS in Eq.~\eqref{eq:Kingbasic} are linear in $\varepsilon_{\mu}$ and $\varepsilon_r$, respectively. Non-linear effects in the King plot could in principle arise at second order in these parameters, with size relative to the leading terms as large as $(A'-A)\varepsilon_{\mu}\gg 10^{-9}$. However, the ratio between the first and second order terms originating from the FS is independent of the transition up to corrections due to the overlap of the electron wave-function with the nucleus, resulting in an extra suppression of $\mathcal{O}(\varepsilon_r)$. Hence, non-linear effects from neglected FS corrections are at most $\mathcal{O}[(A'-A)^2\varepsilon_\mu^2)]\sim 10^{-14}$. There are other neglected effects from the specific MS. The leading contribution to the specific MS is $\mathcal{O}(m_e\mu_{AA'})$, while the sub-leading terms are $\mathcal{O}[\alpha^2 m_e^2(1/m_{A'}^2-1/m_A^2)]$~\cite{Palmer}. Therefore, non-linear effects from the MS are  $\mathcal{O}[\alpha^2m_e(m_{A'}+m_A)/(m_Am_{A'})]\sim 10^{-10}$ level, which is small enough especially since the MS is typically sub-dominant to the FS for heavy nuclei~\cite{ISKing}.  The simple argument above suggests that the breaking of King's linearity from residual QED and nuclear corrections is negligible. A more rigorous check of the negligibility of residual QED corrections should be performed using advanced atomic structure many-body, relativistic, calculations. A first attempt in this direction~\cite{Flambaum:2017onb} recently evaluated the size non-linearities in the E2 and E3 transitions of Yb$^+$ to be $\mathcal{O}(2\,$kHz), which is larger than the above estimate by several orders of magnitude. This difference originates from an enhancement of the quadratic FS contribution due to many-body effects that happen to be particularly strong in the presence of several valence electrons.

{\bf Discussion}
As a proof of concept, one can use the existing IS measurements in Ca$^+$ ($Z=20$) for transitions involving the $4S$ state~\cite{isotope_shift_measurement}. With an error of $\mathcal{O}(100)$kHz and assuming $S_{21}\sim\mathcal{O}(1)$, this results in a rather weak bound of $y_e y_n\lesssim 120$. However, for light mediator we expect a much larger effect and this measurement might already lead to a meaningful bound~\mbox{
\cite{Berengut:2017zuo}}. 
In fact, there are several well motivated examples where the Higgs mixes with a light scalar that inherits its couplings to fermions from the Higgs and thus is effectively described by the above formalism but without a suppressed  $S_{12}$. This was analyzed in~\cite{Frugiuele:2016rii} which found that in the future this limit would be able to probe unprecedented regions of the parameter space, especially in cases where the Higgs coupling to the charge leptons is enhanced relative to the Standard Model~\cite{Flacke:2016szy,Batell:2016ove,Giudice:2008uua,hierarchion}.
 
As another example for a system that could be used to probe new interactions, consider the radio-frequency E1 transitions in Dy atoms. While measurements have already been performed with very high accuracy~\cite{PhysRevA.50.132}, there is very good prospects for significant improvements~\cite{Budker:private}. A careful analysis of the electronic levels of Dy is however required in order to determine whether these systems are suitable for this purpose.
Another possibility is to compare IS for clock transitions in an ion and its corresponding neutral atom, as done {\it e.g.} for  Yb and Yb$^+$~\cite{PhysRevA.85.012502}. Since the nuclear parameters are the same for the ion and the neutral atom, the above analysis still holds. Therefore, additional Higgs-like forces can also be probed with non-linear King plots (as described by Eq.~\eqref{eq:KingCool}) beyond the Yb$^+$ case, using other systems like Ca, Sr and Hg, all of which have narrow clock transitions for the ion and the atom and at least four stable isotopes~\cite{2014arXiv1401.2378P}.  Moreover, many-body calculations recently estimated the breaking of King's linearity in Ca$^+$ and Sr$^+$ to be $\mathcal{O}(1\,$Hz)~\cite{Flambaum:2017onb}, hence much smaller than in Yb$+$. We emphasise that this method can be rather effective in bounding new forces coupled to electrons and neutrons and whose range is comparable or longer than the typical nucleus size. It could then  lead to stringent bounds on the presence of light bosonic mediators~\cite{Berengut:2017zuo}.\\

{\it\bf Acknowledgments:} We thank D.~Budker, V.~Flambaum, C.~Frugiuele, E.~Fuchs, C.~Grojean, K.~Pachucki, G.~Paz, M.~Pospelov and M.~Schlaffer for discussions and comments on the manuscript.  We are especially grateful to D.~Budker,  V.~Flambaum, K.~Pachucki, M.~Pospelov for insightful correspondence.
The work of CD is supported by the ``Investissements d'avenir, Labex ENIGMASS''. 
The work of RO is supported by grants from ISF, ERC, I-CORE, IMOS, and the Crown Photonics Center.
The work of GP is supported by grants from the BSF, ISF, ERC and UK-Weizmann making connections. 
The work of YS is supported by the U.S. Department of Energy under grant Contract Number  DE-SC0012567.

\bibliographystyle{apsrev}
\bibliography{AtomicHiggs-bib}

\end{document}